\documentclass[12pt]{article}
\usepackage{a4wide,epsfig,psfrag,amsmath,amssymb,cite,scalefnt}

\newcommand{\gsim}{\;\rlap{\lower 3.5 pt \hbox{$\mathchar \sim$}} \raise 1pt
 \hbox {$>$}\;}
\newcommand{\lsim}{\;\rlap{\lower 3.5 pt \hbox{$\mathchar \sim$}} \raise 1pt
 \hbox {$<$}\;}

\begin{document}

\title{\vskip-3cm{\baselineskip14pt
    \begin{flushleft}
      \normalsize SFB/CPP-08-56\\
      \normalsize TTP08-34
  \end{flushleft}}
  \vskip1.5cm
  Fermionic contributions to the\\ three-loop static potential
}
\author{\small Alexander V. Smirnov$^{(a,c)}$, 
  Vladimir A. Smirnov$^{(b,c)}$, 
  Matthias Steinhauser$^{(c)}$\\[1em]
  {\small\it (a) Scientific Research Computing Center,
    Moscow State University}\\
  {\small\it 119992 Moscow, Russia}\\
  {\small\it (b) Nuclear Physics Institute,
    Moscow State University}\\
  {\small\it 119992 Moscow, Russia}\\
  {\small\it (c) Institut f{\"u}r Theoretische Teilchenphysik,
    Universit{\"a}t Karlsruhe (TH)}\\
  {\small\it Karlsruhe Institute of Technology (KIT)}\\
  {\small\it 76128 Karlsruhe, Germany}
}

\date{}

\maketitle

\thispagestyle{empty}

\begin{abstract}
We consider the three-loop corrections to the static potential
which are induced by a closed fermion loop. For the reduction of the occurring
integrals a combination of the Gr\"obner and
Laporta algorithm has been used and the evaluation of the master integrals has
been performed with the help of the Mellin-Barnes technique.
The fermionic three-loop corrections amount to 2\% of the tree-level
result for top quarks, 8\% for bottom quarks and
27\% for the charm quark system.

\medskip


\end{abstract}

\thispagestyle{empty}


\newpage


\section{Introduction}

The potential between a heavy quark and its anti-quark is a crucial
quantity both for understanding fundamental properties of QCD, such as
confinement, and for describing the rich phenomenology of heavy
quarkonia~\cite{Brambilla:2004wf} (see also Ref.~\cite{Vairo:2007id} for a
recent review about the static potential).

Within perturbation theory the static potential can be computed 
as an expansion in the strong coupling $\alpha_s$ and the inverse
heavy-quark mass or, equivalently, in the heavy-quark velocity $v$. 
The leading order result in $v$ is known up to the two-loop
approximation~\cite{Fischler:1977yf,Billoire:1979ih,Peter:1996ig,Peter:1997me,Schroder:1998vy,Kniehl:2001ju}
which has been completed about ten years ago. 
For the three-loop corrections there are only estimates relying
on Pad\'e approximations~\cite{Chishtie:2001mf}
or based on renormalon studies~\cite{Pineda:2001zq}.

A new feature of the three-loop corrections is the appearance of 
an infrared divergence which was discussed for the first time in
Ref.~\cite{Appelquist:1977es}. A quantitative analysis of this effect can be
found in Ref.~\cite{Brambilla:1999qa} (see also Ref.~\cite{Brambilla:1999xf})
where a proper definition of the static
potential within perturbation theory is provided. Furthermore, it is argued
that the infrared singularities cancel in physical quantities after including
the contribution where so-called ultra-soft gluons interact with the 
heavy quark anti-quark bound state (see also, e.g.,
Refs.~\cite{Kniehl:1999ud,Kniehl:2002br,Penin:2002zv}). 
Higher order logarithmic contributions to the infrared behaviour of the static
potential have been considered in Refs.~\cite{Pineda:2000gza,Brambilla:2006wp}.

In this paper we compute the fermionic contribution to the three-loop static
potential which is infrared safe. Partial results have already been published
in Refs.~\cite{Smirnov:2008tz,Smirnov:2008ay}.

The static potential enters as a building block in a variety of
physical quantities. Often at three-loop order only estimations are
used or the three-loop coefficient --- usually called $a_3$ ---
appears as a parameter in the final 
result. Let us in this context mention the determination of the bottom
and top quark mass from the ground state energy of the heavy quark
system which has been computed to third order in Ref.~\cite{Penin:2002zv}.
The error on the mass values due to the unknown three-loop coefficient
amounts to 14\% (13\%) of the total uncertainty for the bottom (top) quark.
Similarly, $a_3$ enters the calculation of the total cross section for
top quark threshold production at
next-to-next-to-next-to leading order (see, e.g., Ref.~\cite{Beneke:2007uf}).
The static energy between two heavy quarks has often been used in
order to compare perturbative calculations with simulations on 
the lattice (see, e.g.,
Refs.~\cite{Bali:1999ai,Necco:2001gh,Pineda:2002se,Sumino:2005cq}). Also in
this context the 
knowledge of $a_3$ is crucial and is expected to lead to a better agreement
between the two approaches~\cite{Pineda:2002se}.
Last not least let us mention the extraction of the strong coupling
from lattice simulations where again the static potential plays a
crucial role~\cite{Davies:2008sw,Maltman:2008bx} and the knowledge of $a_3$
would be highly desirable. 

Let us for completeness mention that the one-loop
mass-suppressed corrections to the static potential have been evaluated in
Refs.~\cite{Gupta:1982qc,Pantaleone:1985uf,Titard:1993nn,Manohar:1997qy,Pineda:1998kj,Manohar:2000hj},
the two-loop terms in Ref.~\cite{Kniehl:2001ju}.
Light quark mass effects have been considered in Ref.~\cite{Melles:2000dq}.
A collection of all relevant contributions needed up to
next-to-next-to-next-to-leading order can be found in
Ref.~\cite{Kniehl:2002br}. 
Furthermore, the two-loop corrections for the case
where the quark and anti-quark form an octet state have been evaluated in
Ref.~\cite{Kniehl:2004rk}.

The remainder of the paper is organized as follows: 
In the next Section we present details of our calculation. In particular
we discuss the various types of Feynman diagrams which occur at three-loop
order and their contributions to the individual colour factors.
In Section~\ref{sec::results} our results are presented and
Section~\ref{sec::concl} contains our conclusions.


\section{Calculation}

\begin{figure}[t]
  \centering
  \leavevmode
  \epsfxsize=\textwidth
  \epsffile[100 330 560 490]{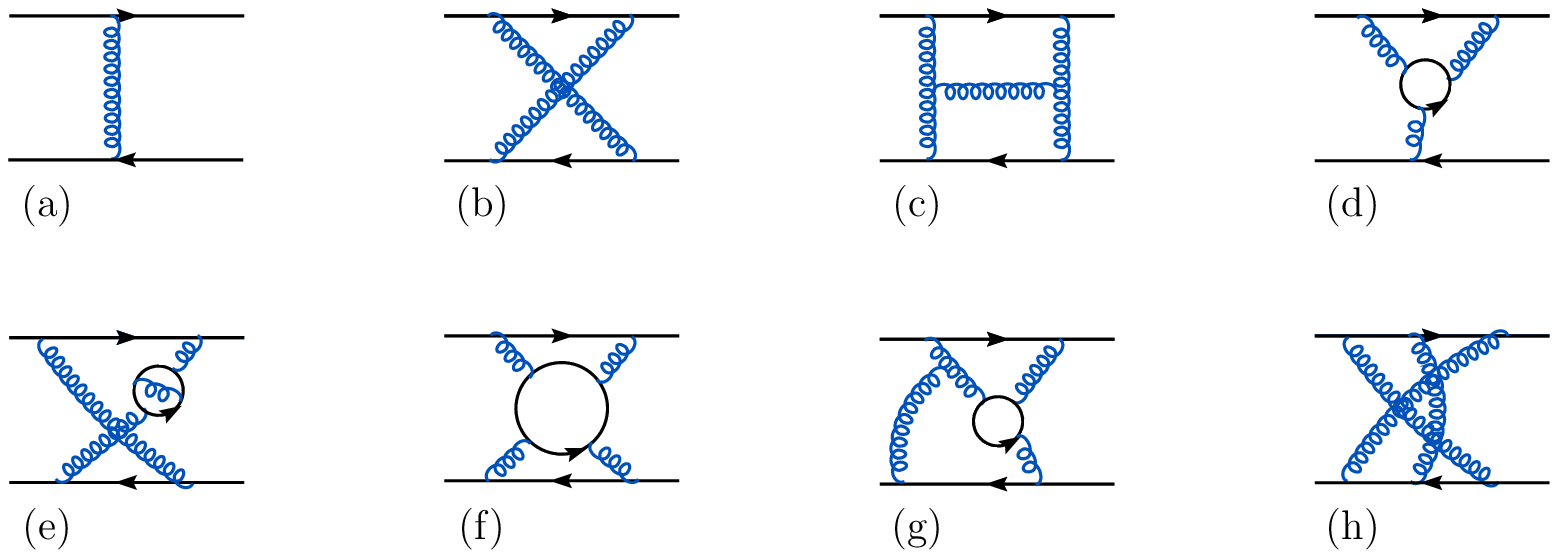}
  \caption{\label{fig::diags}Sample diagrams contributing to the
    static potential at tree-level, one-, two- and three-loop order.
    In this paper only the fermionic corrections are considered at three-loop
    order which excludes diagrams of type (h).
    }
\end{figure}

In the practical calculation of quantum corrections to the static potential
one has to consider a heavy quark and its
anti-quark which interact via the exchange of gluons. In Fig.~\ref{fig::diags}
some sample diagrams up to three-loop order are shown.

In momentum space the static potential can be cast into the form 
\begin{eqnarray}
  V(|{\vec q}\,|)&=&-{4\pi C_F\alpha_s(|{\vec q}\,|)\over{\vec q}\,^2}
  \Bigg[1+{\alpha_s(|{\vec q}\,|)\over 4\pi}a_1
    +\left({\alpha_s(|{\vec q}\,|)\over 4\pi}\right)^2a_2
    \nonumber\\&&\mbox{}
    +\left({\alpha_s(|{\vec q}\,|)\over 4\pi}\right)^3
    \left(a_3+ 8\pi^2 C_A^3\ln{\mu^2\over{\vec q}\,^2}\right)
    +\cdots\Bigg]\,,
  \label{eq::V}
\end{eqnarray}
where $\alpha_s$ denotes the strong coupling in the $\overline{\rm MS}$ scheme and
explicit results for 
$a_1$~\cite{Fischler:1977yf,Billoire:1979ih} 
and $a_2$~\cite{Peter:1996ig,Peter:1997me,Schroder:1998vy,Kniehl:2001ju} 
are given below in Eq.~(\ref{eq::a1a2}). The infrared logarithm at order
$\alpha_s^3$ follows the conventions of Ref.~\cite{Kniehl:2002br} and
the renormalization group logarithms $\ln(\mu^2/{\vec q}\,^2)$ 
can be recovered with the help of
\begin{eqnarray}
  \frac{\alpha_s(|{\vec q}\,|)}{\pi}&=&
  \frac{\alpha_s(\mu)}{\pi}\left[1+\frac{\alpha_s(\mu)}{\pi}\beta_0L
    +\left(\frac{\alpha_s(\mu)}{\pi}\right)^2L\left(\beta_0^2L+\beta_1\right)
  \right.
  \nonumber\\
  &&{}+\left.\left(\frac{\alpha_s(\mu)}{\pi}\right)^3L
    \left(\beta_0^3L^2+\frac{5}{2}\beta_0\beta_1L+\beta_2\right)+\cdots\right]
  \,,
\end{eqnarray}
where $L=\ln(\mu^2/{\vec q}\,^2)$ and the coefficients of the $\beta$ function
(see, e.g., Ref.~\cite{Chetyrkin:2004mf}) read
\begin{eqnarray}
  \beta_0&=&{1\over4}\left({11\over3}C_A-{4\over3}T_Fn_l\right)\,,
  \nonumber\\
  \beta_1&=&{1\over16}\left({34\over3}C_A^2-{20\over3}C_AT_Fn_l-4C_FT_Fn_l
  \right)\,,
  \nonumber\\
  \beta_2&=&{1\over64}\left({2857\over54}C_A^3-{1415\over27}C_A^2T_Fn_l
    -{205\over9}C_AC_FT_Fn_l+2C_F^2T_Fn_l+{158\over27}C_AT_F^2n_l^2
  \right.\nonumber\\
  &&\mbox{}+\left.{44\over9}C_FT_F^2n_l^2\right)
  \,.
\end{eqnarray}
Here, $C_A=N_c$ and $C_F=(N_c^2-1)/(2N_c)$ 
are the eigenvalues of the quadratic Casimir
operators of the adjoint and fundamental representations of the 
$SU(N_c)$ colour gauge group, respectively,
$T_F=1/2$ is the index of the fundamental representation, and $n_l$ is
the number of light-quark flavours.
Let us at this point only mention that $a_1$ has the colour structures
$C_A$ and $T_F n_l$ where the latter contribution originates from one-loop
fermionic corrections to the gluon propagator. Note that there is no colour
factor $C_F$ since this contribution is generated via an iteration of
the tree-level result. 
Consider, e.g., the one-loop planar-ladder and the crossed-ladder diagram of
Fig.~\ref{fig::diags}(b) with the colour factors $C_F^2$ and $C_F^2-C_AC_F/2$,
respectively. It is easy to see (see, e.g., Ref.~\cite{Peter-diss}) that the
$C_F^2$ term can be generated by iterations of the leading order diagram in
Fig.~\ref{fig::diags}(a) leaving only the $C_A C_F$ term as genuine one-loop
contribution.

Similarly, at two-loop order there are the colour factors 
$C_A^2$, $C_A T_F n_l$,  $C_F T_F n_l$, and $(T_F n_l)^2$ where the
latter two originate from loop corrections to
the gluon propagator connecting the quark and the anti-quark.
$a_2$ contains no $C_F^2$ and $C_F C_A$ terms since
their contribution is again generated by iterations of lower-order
results. 

The rule that a colour factor $C_F$ can only arise from corrections to a
fermion bubble in a gluon line also holds at three-loop level. This 
requires a careful analysis of the colour factors for each class of diagram. 
For example, the colour factor of the graph in Fig.~\ref{fig::diags}(e) receives a
contribution\footnote{In addition to the factor $C_F$ already present in
  Eq.~(\ref{eq::V}).}
$(C_F-C_A/2)T_F n_l$ from two-loop fermionic subdiagram and a factor
$(C_F-C_A/2)$ from the remaining crossed box structure. Whereas the complete
first factor has to be taken into account only the $C_A$ term of the second
factor contributes to the potential.
In a similar manner all diagrams have to be analyzed
which leads to the colour structures
$C_A^3$, $C_A^2 T_F n_l$, $C_A C_F T_F n_l$,  $C_F^2 T_F n_l$,
$C_A (T_F n_l)^2$, $C_F (T_F n_l)^2$ and $(T_F n_l)^3$. In this paper we
compute all coefficients except the one of $C_A^3$.
The results for the structures $C_A (T_F n_l)^2$, $C_F (T_F n_l)^2$ and 
$(T_F n_l)^3$ can be found in Refs.~\cite{Smirnov:2008tz,Smirnov:2008ay}.

At three-loop order there is a new class of diagrams containing a
``light-by-light'' subdiagram (see Fig.~\ref{fig::diags}(f) for a sample
graph) which develops the colour factors $d_F^{abcd}d_F^{abcd}/N_A$ and
$C_A^2 T_F n_l$. Note that these contributions are not connected to iterations
and are thus already present in QED (i.e. for $C_A=0$, $d_F^{abcd}=1$, $N_A=1$
and $T_F=1$).

\begin{figure}[t]
  \centering
  \leavevmode
  \epsfxsize=\textwidth
  \epsffile[100 330 560 490]{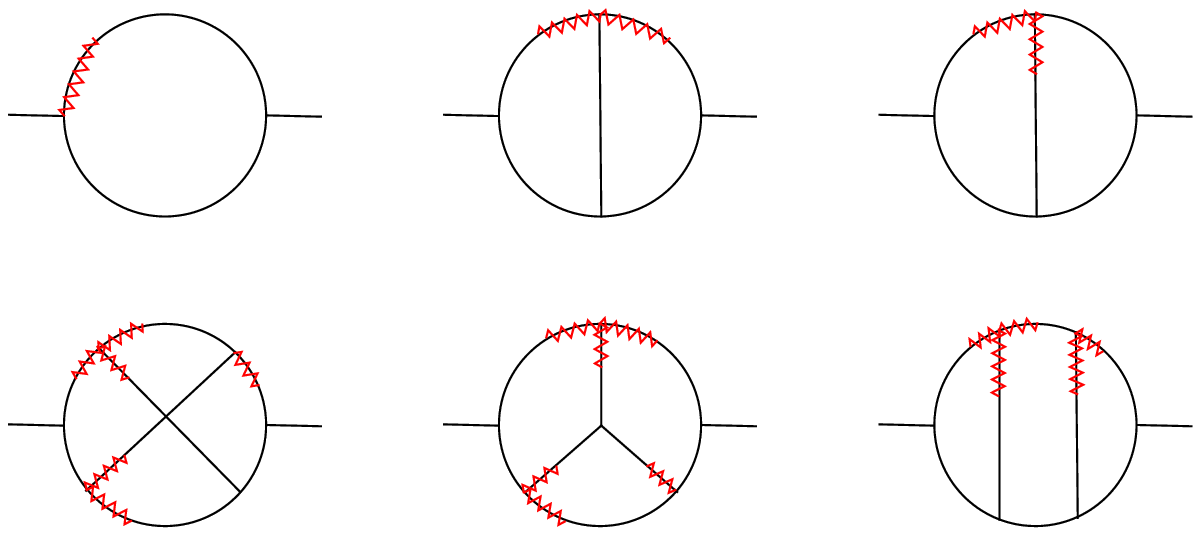}
  \caption{\label{fig::scalar}One-, two- and three-loop diagrams.
    The solid line stands for massless relativistic propagators and the 
    zigzag line represents static propagators.
    }
\end{figure}

Since we use non-relativistic QCD as a starting point for the evaluation of the Feynman
diagrams the momentum transfer between the quark and the anti-quark represents
the only relevant scale in the problem. Thus all integrals can be mapped to
the two-point functions which are shown in Fig.~\ref{fig::scalar} in
diagrammatical form. Next to purely massless lines originating from the gluon,
ghost and light-quark propagators also static lines from the heavy quarks are
present.
The one- and two-loop diagrams have been extensively 
studied in Ref.~\cite{Peter-diss,Schroder:1999sg,Smirnov:2003kc}. As far as
the three-loop diagrams 
are concerned one can perform a partial fractioning in those cases
where three static lines meet at a vertex. This leads to many different 
three-loop graphs involving, however, at most three static lines. 
Thus any resulting integral is labeled by twelve indices one of which
corresponds to an irreducible numerator.

\begin{figure}[t]
  \centering
  \leavevmode
  \epsfxsize=\textwidth
  \epsffile[70 340 580 470]{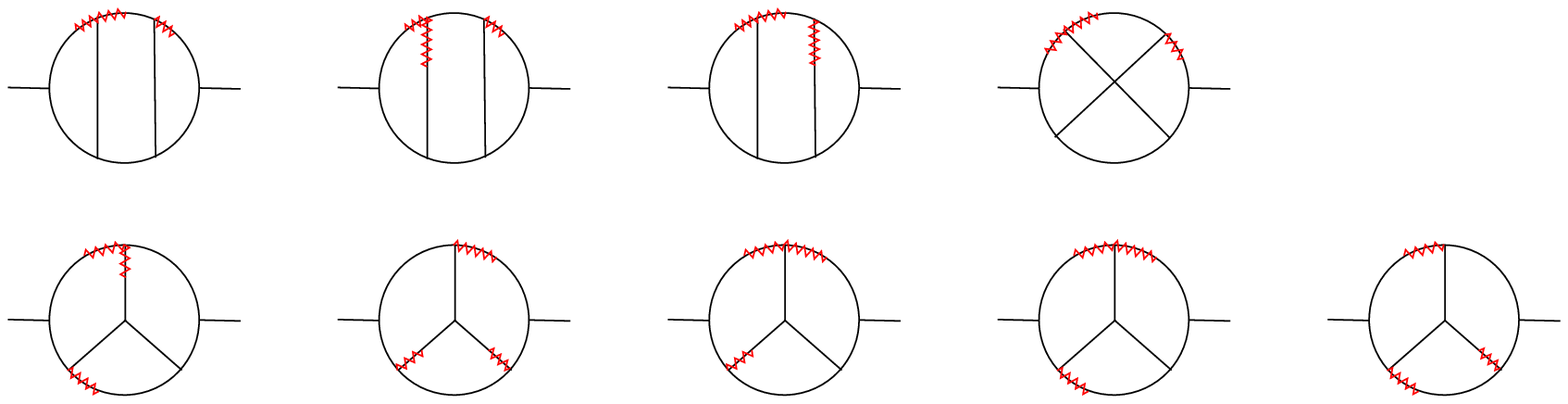}
  \caption{\label{fig::graphs}Three-loop diagrams of ``ladder'' (first three
    diagrams), ``non-planar'' (forth diagram in upper row) and ``mercedes'' 
    type (lower row) which have to be considered for the fermionic part of $a_3$.
    }
\end{figure}

Altogether we have to consider about 70\,000 integrals
(allowing for a general QCD gauge parameter $\xi$)
which can all be mapped to one of the diagrams shown 
in Fig.~\ref{fig::graphs}.
Thereby the linear propagators can appear in two variants: either in the
form $(-v\cdot k-i0)$ or $(-v\cdot k+i0)$. 
If the loop momenta, $k$, $l$ and $r$, in the upper row of 
Fig.~\ref{fig::graphs} are chosen as the momenta of the three
upper lines, then the first diagram appears in two ways: 
either with the product
\[(-v\cdot k-i0)^{-a_9}(-v\cdot l-i0)^{-a_{10}} (-v\cdot r-i0)^{-a_{11}},\]
or with the product
\[(-v\cdot k+i0)^{-a_9}(-v\cdot l-i0)^{-a_{10}} (-v\cdot r-i0)^{-a_{11}}.\]
The second diagram appears with similar propagators where
the momenta $\{k,l,r\}$ (in the first variant with $-i0$ in all three terms)
are replaced by $\{k,k-l,r\}$.
The third diagram in the upper row of Fig.~\ref{fig::graphs} 
corresponds to $\{k,l,l-r\}$ and the fourth one to $\{k,l,r\}$.
In the case of the ``mercedes'' type diagrams in the lower row of
Fig.~\ref{fig::graphs} one  chooses the loop momenta
$k$, $l$ and $r$ as the momenta of the three lower lines. Then
the five diagrams appear with static propagators of the form
$(-v\cdot k-i0)$
with momenta $\{k,k-r,l\}$, $\{r,k-l,r-l\}$, $\{k,r,k-l\}$, $\{k,r,l\}$,
$\{k,r-l,l\}$, respectively.

For the calculation of the diagrams we proceed in the following way: 
They are generated with {\tt
  QGRAF}~\cite{Nogueira:1991ex} and further processed with {\tt q2e} and {\tt
  exp}~\cite{Harlander:1997zb,Seidensticker:1999bb} where a mapping to the
diagrams of Fig.~\ref{fig::scalar} is achieved. In a next step the reduction
of the integrals is performed with the program package {\tt FIRE}~\cite{FIRE} 
which implements a combination of the Laporta~\cite{Laporta:1996mq}
and the Gr\"obner algorithm (see, e.g., Ref.~\cite{Buch}). 
This leads us to about 100 master
integrals which have to be evaluated in an expansion in $\epsilon$
with the help of the Mellin-Barnes technique.
Non trivial examples are discussed in
Refs.~\cite{Smirnov:2008tz,Smirnov:2008ay} where also explicit results
are given. 
We managed to compute all but four 
coefficients of the $\epsilon$ expansion analytically.
As a crucial tool providing very important numerical cross checks of the
analytical results we applied the program {\tt FIESTA}~\cite{FIESTA}
which is a convenient and efficient implementation of the sector
decomposition algorithm.
Finally, let us mention that we evaluate the colour factors with the
help of the program {\tt color}~\cite{vanRitbergen:1998pn}. 

In our calculation we allowed for a general gauge parameter $\xi$ in the
gluon propagator and checked that $\xi$ drops out in the final result.
This constitutes a strong check on the correctness of our result.

In order to obtain a finite result one has to renormalize the strong coupling
which we perform within the $\overline{\rm MS}$ scheme. The corresponding 
renormalization constant can, e.g., be found in Ref.~\cite{Chetyrkin:2004mf}.


\section{\label{sec::results}Results}

Let us in a first step present the results for the one- and two-loop 
coefficients including higher orders in $\epsilon$ since these terms are
needed for the renormalization procedure. We obtain
\begin{eqnarray}
  a_1 &=& {31\over 9}C_A-{20\over 9}T_Fn_l
  + \epsilon\Bigg[ 
    \left(\frac{188}{27} - \frac{11\pi^2}{36}\right)C_A 
    + \left(-\frac{112}{27} + \frac{\pi^2}{9}\right)T_F n_l
    \Bigg]
  \nonumber\\&&\mbox{}
  + \epsilon^2\Bigg[ 
    \left(\frac{1132}{81} - \frac{31\pi^2}{108} 
    - \frac{77\zeta(3)}{9}\right) C_A
    + \left(-\frac{656}{81} + \frac{5\pi^2}{27} 
    + \frac{28\zeta(3)}{9}\right) T_F n_l
    \Bigg]
  \,,\nonumber\\
  a_2 &=&
  \left({4343\over162}+4\pi^2-{\pi^4\over4}+{22\over3}\zeta(3)\right)C_A^2
  -\left({1798\over81}+{56\over3}\zeta(3)\right)C_AT_Fn_l
  \nonumber\\
  &&{}-\left({55\over3}-16\zeta(3)\right)C_FT_Fn_l
  +\left({20\over9}T_Fn_l\right)^2
  + \epsilon\Bigg[ 
      \left(\frac{51637}{972} 
    - \frac{1759\pi^2}{81} 
    + \frac{31\pi^4}{10} 
    \right.\nonumber\\&&\left.\mbox{}
    - \frac{220\zeta(3)}{9} 
    - \pi^4\ln2\right)C_A^2
    + \left(-\frac{11665}{243} 
    + \frac{217\pi^2}{81} 
    - \frac{2\pi^4}{3} 
    - \frac{4\zeta(3)}{9}\right) C_A T_F n_l
  \nonumber\\&&\mbox{}
    + \left(-\frac{1711}{18} 
    + \frac{\pi^2}{3} 
    + \frac{4\pi^4}{15} 
    + \frac{152\zeta(3)}{3}\right) C_F T_F n_l
    + \left(\frac{4480}{243} - \frac{40\pi^2}{81}\right) T_F^2 n_l^2
  \Bigg]
  \,,
  \label{eq::a1a2}
\end{eqnarray}
where the one- and two-loop results (in the limit $\epsilon\to0$) can be found
in Refs.~\cite{Fischler:1977yf,Billoire:1979ih}
and~\cite{Peter:1996ig,Peter:1997me,Schroder:1998vy,Kniehl:2001ju},
respectively.
In Eq.~(\ref{eq::a1a2}) $\zeta$ is Riemann's zeta function, with the value
$\zeta(3)=1.202057\ldots$.

The three-loop result can be cast in the form
\begin{eqnarray}
  a_3 &=& a_3^{(3)} n_l^3 + a_3^{(2)} n_l^2 + a_3^{(1)} n_l + a_3^{(0)}
  \,,
\end{eqnarray}
where the first three coefficients on the right-hand side read
\begin{eqnarray}
  a_3^{(3)} &=& - \left(\frac{20}{9}\right)^3 T_F^3
  \,,\nonumber\\
  a_3^{(2)} &=&
  \left(\frac{12541}{243}
    + \frac{368\zeta(3)}{3}
    + \frac{64\pi^4}{135}
  \right) C_A T_F^2
  +
  \left(\frac{14002}{81}
    - \frac{416\zeta(3)}{3}
  \right) C_F T_F^2
  \,,\nonumber\\
  a_3^{(1)} &=&
  \left(-709.717
  \right) C_A^2 T_F
  +
  \left(-\frac{71281}{162}
    + 264 \zeta(3)
    + 80 \zeta(5)
  \right) C_AC_F T_F
  \nonumber\\&&\mbox{}
  +
  \left(\frac{286}{9}
    + \frac{296\zeta(3)}{3}
    - 160\zeta(5)
  \right) C_F^2 T_F
  +
  \left(-56.83(1)
  \right) \frac{d_F^{abcd}d_F^{abcd}}{N_A}
  \,.
  \label{eq::a3}
\end{eqnarray}
where the $SU(N_c)$ colour factors are given by
\begin{eqnarray}
  &&C_A = N_c\,,\quad C_F=\frac{N_c^2-1}{2 N_c}\,,\quad T_F=\frac{1}{2}\,,
\quad
  \frac{d_F^{abcd}d_F^{abcd}}{N_A} = \frac{18 - 6 N_c^2 + N_c^4}{96 N_c^2}
  \,.
\end{eqnarray}
In Eq.~(\ref{eq::a3}) only the coefficient of $d_F^{abcd}d_F^{abcd}$ is
affected by the limited numerical precision of the four coefficients 
only known numerically which is indicated by the number in
round brackets.

We are now in the position to briefly discuss the numerical effect of
the new corrections. Inserting the results for $a_1$, $a_2$ and $a_3$ 
in Eq.~(\ref{eq::V}) it takes the form
\begin{eqnarray}
  V(|{\vec q}\,|)&=&-{4\pi C_F\alpha_s(|{\vec q}\,|)\over{\vec q}\,^2}
  \Bigg[1+\frac{\alpha_s}{\pi}\left(2.5833 - 0.2778 n_l\right)
    \nonumber\\&&\mbox{}
    +\left(\frac{\alpha_s}{\pi}\right)^2\left(28.5468 - 4.1471 n_l 
    + 0.0772 n_l^2 \right)
    \nonumber\\&&\mbox{}
    +\left(\frac{\alpha_s}{\pi}\right)^3\left(
    \frac{a_3^{(0)}}{4^3} -51.4048 n_l 
    + 2.9061 n_l^2  - 0.0214 n_l^3\right)
    +\cdots\Bigg]\,,
  \label{eq::Vnum}
\end{eqnarray}
where the ellipses denote higher order terms and $\mu^2={\vec q}\,^2$ has
been chosen in order to suppress the infrared logarithm.
From Eq.~(\ref{eq::Vnum}) one observes that both at one- and two-loop
order the linear $n_l$ term is negative and leads to a screening of the
(positive) non-$n_l$ contribution by an amount of about 50\% for
$n_l=5$. Also at three-loop order the linear $n_l$ term is negative
and has a sizeable coefficient. Both for $a_2$ and
$a_3$ the $n_l^2$ contribution is small; the same is true for the 
$n_l^3$ term of $a_3$.

\begin{table}[t]
  \begin{center}
    \begin{tabular}{c|l|l|l|l}
      $n_l$ & $\alpha_s^{(n_l)}$ & 1 loop & 2 loop & 3 loop \\
      \hline
      3 & 0.40 & 0.2228 & 0.2723 
      & $32.25\,\cdot\,10^{-6} a_3^{(0)} - 0.2655$ \\
      4 & 0.25 & 0.1172 & 0.08354 
      & $7.874\,\cdot\,10^{-6} a_3^{(0)}- 0.08088$\\
      5 & 0.15 & 0.05703 & 0.02220 
      & $1.701\,\cdot\,10^{-6} a_3^{(0)} - 0.02036$
    \end{tabular}
    \caption{\label{tab::a123}Radiative corrections to the potential
      $V(|{\vec q}\,|)$ where the tree-level result is normalized to 1
      (cf. Eq.~(\ref{eq::Vnum})).}
  \end{center}
\end{table}

In Tab.~\ref{tab::a123} we present the one-, two- and three-loop
results from the square bracket of Eq.~(\ref{eq::Vnum}) and choose
$n_l$ according to the charm, bottom and top quark
case. In the second column we also provide the numerical value of
$\alpha_s$ corresponding to the soft scale where $\mu\approx m_q\alpha_s$
(and $m_q$ is the heavy quark mass).
It is interesting to note that the three-loop corrections computed in
this paper lead to corrections which are of the same order of
magnitude as the two-loop corrections, however, with a different sign.
In fact, one obtains corrections of about $-27\%$, $-8\%$ and $-2\%$
for charm, bottom and top quarks, respectively.
Furthermore, let us mention that the unknown constant has to be of the
order $10^4$ (and positive) in order to significantly reduce the size
of the three-loop corrections.

Let us now compare our explicit calculation with the
predictions based on Pad\'e approximation. In
Refs.~\cite{Chishtie:2001mf} and~\cite{Pineda:2001zq}
one can find for $a_3/4^3$ the results
$\{313,250,193,142,97.5,60.1,30.5\}$ and
$\{292,227,168,116,72,37,12\}$, respectively, 
where the entries in the list
correspond to $n_l=0,\ldots,n_l=6$. A fit to a cubic polynomial in
$n_l$ leads to 
$a_3/4^3 \approx 380.9 - 70.42 n_l + 2.34 n_l^2  + 0.08 n_l^3$ and
$a_3/4^3 \approx 362.0 - 72.17 n_l + 2.00 n_l^2  + 0.17 n_l^3$,
respectively.
The comparison to Eq.~(\ref{eq::Vnum}) shows that the coefficients 
have the correct sign (except the one of $n_l^3$ which is, however,
close to zero) and the correct order of magnitude. 
Let us nevertheless mention that the (numerically big)
coefficient of the linear $n_l$ term 
deviates by about 50\%.

Finally we want to specify our result for $V(|{\vec q}\,|)$ to QED
which describes the potential of two heavy leptons in the
presence of $n_l$ massless leptons. Substituting for the colour
factors $C_A=0$, $C_F=1$, $T_F=1$, $d_F^{abcd}=1$ and $N_A=1$
we obtain
\begin{eqnarray}
  V_{\rm QED}(|{\vec q}\,|)&=&
  -{4\pi \bar{\alpha}\over{\vec q}\,^2}
  \Bigg[1 + \frac{\bar{\alpha}}{\pi} \left(- 0.5556 n_l \right)
    +\left(\frac{\bar{\alpha}}{\pi}\right)^2\left(0.05622 n_l 
    + 0.3086 n_l^2 \right)
    \nonumber\\&&\mbox{}
    +\left(\frac{\bar{\alpha}}{\pi}\right)^3\left(
     -1.131 n_l + 0.09655 n_l^2  - 0.1715 n_l^3\right)
    +\cdots\Bigg]
  \nonumber\\&=&
  -{4\pi \bar{\alpha}\over{\vec q}\,^2}
  \Bigg[1 - 0.5556\, \frac{\bar{\alpha}}{\pi} 
    + 0.3649\,\left(\frac{\bar{\alpha}}{\pi}\right)^2
    - 1.206 \,\left(\frac{\bar{\alpha}}{\pi}\right)^3
    +\cdots\Bigg]\,,
  \label{eq::VQEDnum}
\end{eqnarray}
where $\bar{\alpha}=\bar{\alpha}(\vec{q}\,^2)$ is the QED coupling in the
$\overline{\rm MS}$ scheme and after the second equality sign $n_l=1$ has been chosen.
This corresponds to a bound state of a muon and an anti-muon in the
presence of a massless electron pair.
The coefficients in Eq.~(\ref{eq::VQEDnum}) are significantly smaller as
in the case of QCD which results in corrections of the order $10^{-8}$
from the three-loop term.

The terms in Eq.~(\ref{eq::VQEDnum}) originate from corrections to the photon
propagator plus the ``light-by-light''-like diagrams as in
Fig.~\ref{fig::diags}(f). Thus for $n_l=1$ $V_{\rm QED}$ can be written in the form
\begin{eqnarray}
  V_{\rm QED}(|{\vec q}\,|)&=&
  -{4\pi\over{\vec q}\,^2}\frac{\alpha}{1+\Pi(\vec{q}\,^2)}
  \Bigg[1 + 
  \left(\frac{\alpha}{\pi}\right)^3 \,n_l\, \left(-0.888\right)
  +\cdots\Bigg]\,,
  \label{eq::VQEDnum2}
\end{eqnarray}
where the photon polarization function is given by
\begin{eqnarray}
  \Pi(\vec{q}\,^2) &=& 
  \frac{\alpha}{\pi}\left(\frac{5}{9} - \frac{L_m}{3}\right) 
  + \left(\frac{\alpha}{\pi}\right)^2 \left(\frac{5}{24} -\zeta(3) -
    \frac{L_m}{4}\right) 
  + \left(\frac{\alpha}{\pi}\right)^3
  \left[-\frac{1703}{1728}  
    - \frac{23}{12}\zeta(2) 
    \right.\nonumber\\&&\left.\mbox{}
    + 2\zeta(2)\ln2
    - \frac{173}{288}\zeta(3)
    + \frac{5}{2}\zeta(5) 
    + \left(\frac{47}{96} - \frac{1}{3}\zeta(3)\right)L_m 
    - \frac{L_m^2}{24} 
  \right]
  \,,
  \label{eq::PiOS}
\end{eqnarray}
with $\zeta(5)=1.036927\ldots$ and $L_m=\ln\vec{q}\,^2/m_e^2$ where $m_e$ is
the electron mass.
In Eqs.~(\ref{eq::VQEDnum2}) and~(\ref{eq::PiOS}) we have used the fine
structure constant $\alpha$. The three-loop relation to $\bar\alpha$ can be
found in Ref.~\cite{Broadhurst:1991fi}.


\section{\label{sec::concl}Conclusion and outlook}

In this letter we report about the calculation of the fermionic
corrections to the static potential of a quark and an anti-quark.
All occurring integrals are reduced to about 100 master integrals with
the help of the program {\tt FIRE}.
The main result can be found in Eqs.~(\ref{eq::a3}) where the
three-loop coefficients are given for each occurring colour structure.
The numerical corrections of the new three-loop terms are quite
sizeable when applied to the system of two charm, bottom or top
quarks. However, for a definite conclusion one has to wait for the 
$n_l$ independent three-loop coefficient $a_3^{(0)}$.

The calculation of $a_3^{(0)}$ is currently in progress. We do not
expect any conceptual problems. However, significantly more Feynman
diagrams contribute which leads to many new graphs in addition 
to those shown in Fig.~\ref{fig::graphs}. As a consequence also more master
integrals have to be evaluated.


\vspace*{2em}
{\large\bf Acknowledgements}

We would like to thank Alexander Penin for many useful discussions and
communications. We also thank York Schr\"oder for many interesting 
discussions at the early stage of the project.
We thank Yuichiro Kiyo, Hans K\"uhn, Alexander Penin and York Schr\"oder for
useful comments to the manuscript.
M.S. thanks Luminita Mihaila, Jan Piclum and Jos Vermaseren
for discussions on colour factors. This work was supported by RFBR, grant
08-02-01451 and by the DFG through SFB/TR~9.



\end{document}